\documentclass[review,preprint]{elsarticle}
\newtheorem{thm}{Theorem}
\newtheorem{lem}[thm]{Lemma}

\newproof{pf}{Proof}
\newdefinition{defi}{Definition}

\usepackage{subfigure}
\usepackage{graphicx}
\usepackage{graphics}
\usepackage{amssymb}
\input{epsf}

\usepackage{algorithm}
\usepackage{algpseudocode}

\begin{document}
\begin{frontmatter}
\title{A simple approximation algorithm for the internal Steiner minimum tree}
\author[bang]{Bang Ye Wu\corref{cor1}}
\ead{bangye@ccu.edu.tw}

\address[bang]{Dept. of Computer Science and Information Engineering,
National Chung Cheng University, ChiaYi, Taiwan 621, R.O.C.}
%\author{Chin-Ping Yang\and \and Bang Ye Wu}
%\cortext[cor1]{corresponding author}

\begin{abstract}
For a metric graph $G=(V,E)$ and $R\subset V$, the internal Steiner minimum tree problem asks for a minimum weight Steiner tree spanning $R$ such that every vertex in $R$ is not a leaf.
This note shows a simple polynomial-time $2\rho$-approximation algorithm, in which $\rho$ is the approximation ratio for the Steiner minimum tree problem. The result improves the approximation ratio $2\rho+1$ in \cite{hsieh13}.
\end{abstract}
\end{frontmatter}

\section{Introduction}

For a simple undirected graph $G=(V,E)$ and $R\subset V$, a \emph{Steiner tree} is a connected and acyclic subgraph of $G$ that spans all the vertices in $R$.
The Steiner Minimum Tree problem (SMT) is a well-known NP-hard problem which involves finding a Steiner tree with minimum total weight.
An undirected graph is a \emph{metric graph} if it is a complete graph with nonnegative edge weights satisfying the triangle inequality.
In \cite{hsieh13}, the following variant of SMT is studied. 
\begin{quote}
{\sc Problem}: {\sc Internal Steiner minimum Tree} (ISMT)\\
{\sc Instance}: A metric graph $G=(V,E)$ and $R\subset V$ such that $|V\setminus R|\geq 2$.\\
{\sc Goal}: Find a minimum weight Steiner tree such that every vertex in $R$ is not a leaf.\\
\end{quote}	
Note that there is no solution if $|V\setminus R|< 2$ since any tree has at least two leaves.
In \cite{hsieh13}, it was shown that ISMT is Max SNP-hard and can be approximated with ratio $(2\rho+1)$ in polynomial time, in which $\rho$ is the approximation ratio for SMT. Currently $\rho=\ln 4+\epsilon < 1.39$ on general metric graphs \cite{byrka10}. 
They also showed a 9/7-approximation algorithm for the case that all edge weights are either 1 or 2. For the motivation and the related works, see \cite{hsieh13}.
It should be noted that there exists a $(\rho+1)$-approximation algorithm for a similar problem, named \emph{selected internal Steiner tree problem} (SIST), in which we are given $R'\subset R\subseteq V$ and asked for a Steiner minimum tree spanning $R$ such that every vertex in $R'$ cannot be a leaf \cite{li10}. Although the approximation ratio is better, an additional constraint $|R\setminus R'|\geq 2$ is required.
But, as pointed out in the very end of \cite{hsieh13}, the approximation algorithm in \cite{li10} can also be modified to solve ISMT with the same ratio $(1+\rho)$ with an additional polynomial factor in time complexity. Therefore, the currently best ratio 
achieved in polynomial time is $(1+\rho)$.
 
There is another $2\rho$-approximation algorithm for SIST in \cite{hsieh07}, which can also be modified to solve ISMT. In this note, we show a very simple polynomial-time $2\rho$-approximation algorithm for ISMT.

\section{Algorithm}

For a graph $G$, $V(G)$ and $E(G)$ denote the vertex and the edge sets, respectively.
An edge between vertices $u$ and $v$ is denoted by $(u,v)$, and its weight is denoted by $w(u,v)$. For a subgraph $T$ of $G$, $w(T)$ denotes the total weight of all edges of $T$.
For a vertex subset $U$, the subgraph of $G$ induced by $U$ is denoted by $G[U]$.
By $SMT(G,R)$, we denote a Steiner minimum tree with instance $(G,R)$. 
A path with end vertices $s$ and $t$ is called an $st$-path.

Let $T^*$ denote an optimal solution of ISMT.
Suppose that $s,t\in V\setminus R$ are two leaves in $T^*$. We construct an $st$-path as follows. The doubling-tree method is similar to the one in \cite{tspp00} for solving the traveling salesperson path problem. 

\begin{quote}
First, construct a $\rho$-approximation $T_1$ for $SMT(G[V\setminus \{s,t\}],R)$. 
Next, find $u_1\in V(T_1)$ closest to $s$, i.e., $w(s,u_1)=\min \{w(s,v)|v\in V(T_1)\}$; and similarly find $u_2\in V(T_1)$ closest to $t$.
Then, we construct $T_2= T_1\cup \{(s,u_1),(t,u_2)\}$.
To form the desired $st$-path, we double the edges of $T_2$ except for those on the unique $st$-path on $T_2$. 
The result is a connected multigraph in which the degrees of all vertices except for $s$ and $t$ are even. Therefore we can find an Eulerian walk between $s$ and $t$ 
and turn it into a Hamiltonian $st$-path $P$ without increasing the weight by shortcutting and applying the triangle inequality.	
\end{quote}

\begin{lem}\label{p1}
There exists an optimal solution $T^*$ in which every leaf is adjacent to a vertex in $R$. Furthermore, if $u$ is a leaf and $(u,v)\in E(T^*)$, then $w(u,v)=\min\{w(u,r)| r\in R\}$.
\end{lem}
\begin{pf}
By definition, a leaf must be in $V\setminus R$.
If a leaf is adjacent to a vertex not in $R$, removing the leaf results in a feasible solution without increasing the weight. The second statement directly comes from the optimality and a metric graph is a complete graph.
\qed
\end{pf}
We now show the approximation ratio.
\begin{lem}\label{ratio}
$w(P)\leq 2\rho w(T^*)$.
\end{lem}
\begin{pf}
Since $s$ and $t$ are leaves on $T^*$, removing them results in a tree $T^{-}$ spanning all  vertices in $R$. Since $T_1$ is a $\rho$-approximation of a Steiner minimum tree spanning $R$, we have that $w(T_1)\leq \rho w(T^-)$.

Furthermore, since $s$ and $t$ are connected to their closest vertices in $V(T_1)$, by Lemma~\ref{p1}, we have that $w(T_2)-w(T_1)\leq w(T^*)-w(T^{-})\leq \rho(w(T^*)-w(T^{-}))$, and therefore 
$w(T_2)\leq \rho w(T^*)$.
Finally, by the construction of $P$, $w(P)\leq 2w(T_2)\leq 2\rho w(T^*)$.
\qed\end{pf}
\begin{thm}
The problem ISMT can be $2\rho$-approximated in polynomial time. 
\end{thm}
\begin{pf}
For each pair $s,t$ of vertices in $V\setminus R$, we construct an $st$-path by the above algorithm and output the best one. By Lemma~\ref{ratio}, the approximation ratio is two. 
For the time complexity, 
since there are $O(|V\setminus R|^2)$ pairs of $s,t$, the time complexity is polynomial.
\qed
\end{pf}

\section*{Acknowledgment}
This work was supported in part by 
NSC 100-2221-E-194-036-MY3 and NSC 101-2221-E-194-025-MY3 from the National Science Council, Taiwan, R.O.C.

\end{document}